\documentclass[fleqn,usenatbib]{mnras}
\usepackage[english]{babel}
\usepackage[T1]{fontenc}
\usepackage{ae,aecompl}
\usepackage{fancyhdr}
\usepackage{amsfonts}
\usepackage{amsmath}
\usepackage{amssymb}
\usepackage{multicol}
\usepackage{layout}
\usepackage{graphicx}
\usepackage{multirow}
\usepackage{color}
\usepackage{multirow}
\usepackage{times}
\usepackage{natbib}
\def\be{\begin{equation}}
\def\ee{\end{equation}}

\usepackage[utf8]{inputenc}
\usepackage{lmodern}

\definecolor{valecol}{rgb}{0,0.5, 1.}

\newif\ifAMStwofonts
\AMStwofontstrue

\graphicspath{{./fig/}}

\title{Can decaying dark matter scenarios alleviate both $H_0$ and $\sigma_8$ tensions? }
\author[Davari and Khosravi]{
	Zahra Davari and Nima Khosravi\\ 
	Department of Physics, Shahid Beheshti University, 1983969411,  Tehran, Iran\\}
\begin{document}
	\label{firstpage}
	
	\maketitle	
	\begin{abstract}

	Current tensions in cosmology, including $H_0$ and $\sigma_8$, provide one of the strong reasons to suspect the existence of physics beyond the standard model of cosmology ($\Lambda$CDM). In this paper, we investigate if there is a relation between these tensions and beyond cold dark matter scenarios. To model non-CDM, we assume a decaying dark matter (DDM) which is unstable and may decays into two daughter particles, a combination of cold dark matter, warm dark matter and dark radiation, to explore a vast era of possibilities. We checked our model against CMB data, and could show that decaying dark matter seems not a promising candidate to address the cosmological tensions. 
	
	\end{abstract}
	
	
\section{Introduction}\label{sec:int}
At the present time, the best-fitting scenario for describing the statistics of the universe on large scales is the standard model of cosmology often known as $\Lambda$CDM.
One of the main pillars in the standard $\Lambda$CDM model is dark matter (DM).
The existence of dark matter in our universe is undoubtedly confirmed by numerous kinds of astrophysical observation on a range of length scales from galaxy rotation curves and gravitational lensing to large scale structure (LSS) and the cosmic microwave background (CMB). Nevertheless, the physical nature of DM particles is unclear and enigma after decades of research, largely since cosmic observations are only sensitive to the gravitational effects of DM rather than the properties of its particles~\citep{Scott:2018adl, Pan-STARRS1:2017jku,Planck:2018lbu}.\\
What we already know about DM in the standard model of cosmology $\Lambda$CDM is that it is responsible for about 85\% of the universe matter content and is  a non-luminous, dark component of matter which must be non-relativistic or cold. This means it has a sufficiently small thermal velocity to address formation of the structures and DM particles must be stable on cosmological time scales.\\
Despite remarkable successes in explaining the large scale structure of the universe, the CDM model is currently facing a number of potentially serious problems on the small scales such as too big to fail~\citep{Purcell:2012kd}, missing satellites~\citep{Moore:1999nt}, and core-cusp problems~\citep{Rubin:1970zza,Gentile:2004tb,vanEymeren:2009gq}, as well as on large scales there are some tensions between parameters inferred from local and global cosmological measurements, most notably the Hubble parameter $H_0$ and $\sigma_8$. Early universe observations such as CMB or baryonic acoustic oscillations (BAO) prefer significantly lower value $ H_0=67.27\pm0.60$ km/s/Mpc~\citep{Aghanim:2018eyx} in comparison to the local universe observations such as the calibration
of the cosmic distance ladder scale through Cepheid stars and supernovae type Ia  which determine
$ H_0=74.03\pm1.42 $km/s/Mpc~\citep{Riess:2019cxk}.
The discrepancy between the two $H_0$ values  reaches $\sim4\sigma $~\citep{Verde:2019ivm,Freedman:2019jwv,DiValentino:2020zio}. On the other hand, recent weak lensing surveys such as the joint analysis of $\rm KIDS1000+BOSS+2dfLenShave $ revealed that the direct measurement of the parameter combination $ S_8 \equiv \sigma_8(\Omega_m/0.3)^{0.5} $ – measuring the amplitude of matter fluctuations on $ 8 h^{-1}$Mpc scales is $\sim 3\sigma$ discrepant with the value reconstructed from cosmic microwave background (CMB) data assuming the $\Lambda$CDM model\citep{Aghanim:2018eyx,Hildebrandt:2018yau,Joudaki:2019pmv,KiDS:2020suj}. These may be taken as a hint that the CDM paradigm, although very simple, is not in fact a complete model.\\
It is an interesting question if both small-scale problems and $\Lambda$CDM tensions can be simultaneously resolved through deviations from the standard $\Lambda$CDM cosmological model such as a modification in the nature of the DM component. 
This idea has been studied in different forms e.g.  partially acoustic dark matter models~\citep{Raveri:2017jto}, dissipative dark matter models~\citep{daSilva:2018dzn}, cannibal dark matter~\citep{Buen-Abad:2018mas}, self-
interacting dark matter (SIDM) models~\citep{Loeb:2010gj,Archidiacono:2019wdp} and non-thermal dark matter~\citep{Alcaniz:2019kah}.\\
Since dark matter does not require to be absolutely stable given the cosmological and astrophysical evidence, only it needs to be very long-lived in other words, its lifetime be much longer than the age of the universe~\citep{Ibarra:2013cra}.
Therefore there is a possibility for unraveling mentioned problems and tensions assuming a moderate amount of DM particle decay~\citep{Flores:1986jn,Doroshkevich:1989bf,Enqvist:2015ara,Poulin:2016nat,Vattis:2019efj,Haridasu:2020xaa,Abdalla:2022yfr}. In the simplest case, it is assumed that dark matter decays into  massless particles or very light states in the dark sector such as relativistic dark radiation (DR)~\citep{Audren:2014bca}, but it has been shown to conflict with Planck 2015 data~\citep{Poulin:2016nat,Chudaykin:2016yfk}, the latest Planck CMB lenses and BAO data~\citep{Bringmann:2018jpr}
and a combination of the Pantheon Sample,  observational
Hubble data(OHD), and BAO data~\citep{Anchordoqui:2022gmw}.\\
In this work, we want to investigate the phenomenology of the Decaying Dark Matter (DDM) scenario for cases that DDM experiences 2-body decays where the decay products are different particles.  We study the dynamics of these models and their impacts on the CMB. 
We consider the different cases that cold unstable parent particles decay into  2-body daughter particles as:
\begin{itemize}
\item [i-] decay into two types of stable cold dark matter with various finite masses ($\rm DDM\rightarrow \rm CDM1+CDM2$).
	\item[ ii-]	decay into cold and warm daughter particles ($\rm DDM\rightarrow \rm CDM+WDM$).
	\item[ iii-] decay into cold dark matter particles and relativistic  massless particles or dark radiation ($\rm DDM\rightarrow \rm CDM+DR$).
	\item [iv-] decay into warm dark matter particles and relativistic  massless particles or dark radiation ($\rm DDM     \rightarrow \rm WDM+DR$).
\end{itemize}
The main scenario that is introduced and examined in this study is the second case(CDM-WDM) and other scenarios have already been studied and will be discussed again here for comparison.	
The paper is structured as follows. In section \ref{sec:Boleq}, we derive Boltzmann equations for each model. We implement the related equations in the publicly available numerical code \texttt{CLASS}\footnote{\label{myfootnote0}\url{https://github.com/lesgourg/class_public}}(the Cosmic Linear Anisotropy Solving System)~\citep{Lesgourgues:2011rh} and useing the code \texttt{MONTEPYTHON-v3}\footnote{\label{myfootnote}\url{https://github.com/baudren/montepython_public}}~\citep{Audren:2012wb,Brinckmann:2018cvx} to perform a Monte Carlo Markov chain (MCMC) analysis with a Metropolis-Hasting algorithm for testing different decaying models against the high- CMB TT, TE, EE +low- TT, EE+lensing data from Planck 2018~\citep{Aghanim:2018eyx} and also we review the basic properties of two-body decays and its cosmological implications in section~\ref{sec-data}. We conclude in section~\ref{concl}.
\section{ Boltzmann equations} \label{sec:Boleq}
Most previous studies have investigated cases that dark matter particles decay to  massless (relativistic) particles and massive particles as cold or warm kinds. Here, we decided to inquire about the decaying dark matter particles that decay into other two massive particles with arbitrary smaller masses.\\
We consider decaying dark matter particles (DDM) with mass M decay into another two species of stable dark matter with a small mass splitting, $DDM  \rightarrow \chi_1 + \chi_2$, with masses $m_1$  and $m_2$.\\ 
As we decide to survey the behavior, the evolution of  energy densities and  density perturbations for mother and daughter particles, it is necessary to obtain a complete set of Boltzmann equations that describes the evolution
of all particle species during the decay. So in the first step, we write the Boltzmann equations for the distribution functions of mother and daughter particles both at zeroth order and first order for all cases.\\	
In this paper, we choose the synchronous gauge of the mother particles to describe cosmological linear perturbations but since the \texttt{CLASS} code is written in both synchronous and Newtonian gauges, we can easily convert their equations to each other~\citep{Ma:1995ey}. In the following, we have used ~\cite{Aoyama:2011ba, Wang:2012eka,Aoyama:2014tga,Audren:2014bca,Abellan:2021bpx} in deriving equations to describe the evolution of the particles.\\
The perturbed line element of the FRW metric in the synchronous gauge is given by
\begin{equation}
ds^2=a(\tau)^2\big\{d\tau^2+\big(\delta_{\rm ij}+h_{\rm ij}({\bf{x}},\tau)\big)dx^idx^j\big\},
\end{equation}
where $a$ is the scale factor and $h_{\rm ij}$ is the metric perturbation in the synchronous gauge. We could rewrite the scalar modes of $h_{ij}$ as a Fourier integral in  k-space Fourier by introducing two fields $h({\bf k},\tau)$ and $\eta({\bf k},\tau)$ as:
\begin{eqnarray}
&&h_{\rm ij}({\bf x},\tau)=\\ \nonumber
&&\int d^3k \left(h({\bf k},\tau){\bf \hat{k}_i \hat{k}_j}+6\eta({\bf k},\tau)({\bf \hat{k}_i \hat{k}_j}-\frac{1}{3}\delta_{\rm ij}) \right)e^{i{\bf k.x}};
\end{eqnarray}
where ${\bf k}\equiv k\hat{{\bf k}}$ is a wave number vector and $\hat{{\bf k}}$ is the unit vector of ${\bf k}$. We remind that $h$ is used to represent the trace of  $h_{\rm ij}$ both in real space and in Fourier space~\citep{Ma:1995ey,Aoyama:2014tga}. In the following, we focus on a single mode of perturbations at mode ${\bf k}$.
Time evolution of a phase-space distribution function of particles, $f({\bf{x, q}},\tau )$, can be written according to the Boltzmann equation as
\begin{equation}\label{fmain}
\frac{\partial f}{\partial \tau}+\frac{dx^i}{d\tau} \frac{\partial f}{\partial x^i}+\frac{dq}{d\tau} \frac{\partial f}{\partial q}+\frac{dn^i}{d\tau} \frac{\partial f}{\partial n^i}=\big(\frac{\partial f}{\partial \tau}\big)_c,
\end{equation}
where $q$ is the comoving momentum, which is related to the physical momentum $p$ by $q = ap$, and $n^i = q^i /q$ are the norm and the direction of $q$, respectively.
For a perturbation mode with ${\bf k}$, the second term of the left hand side can be rewritten as $i\frac{q}{\epsilon}({\bf k.\hat{n}})$, here $ \epsilon=\sqrt{q^2+a^2m^2}$ is the comoving energy of particles.
On the left hand side, the third term can be rewritten in terms of the metric perturbations by adopting the geodesic
equation $p^0\frac{dp^\mu}{d\tau}+\Gamma^{\mu}_{\alpha\beta}p^{\alpha}p^{\beta}=0$ for the time component ($\mu=0$), we can rewrite it in the following form:
\begin{equation}
\frac{dq}{d\tau}= \dot{\eta}-\frac{1}{2}(\dot{h}+6\dot{\eta}({\bf k .\hat{n}})^2,
\end{equation}
where dot is a derivative with respect to conformal time. The fourth term is a second order term since both ($dn_i /d\tau$ ) and ($\partial f /\partial n_i$ ) are first-order quantities and can be neglected.
The right hand side of the above equations is the collision term, which also represents effects of decay or creation of particles on the distribution function which vanishes, ($\partial f /\partial \tau)_c = 0$, in the absence of non-gravitational interactions. So we can rewrite the equation \eqref{fmain} as
\begin{equation}\label{fmain2}
\frac{\partial f}{\partial \tau}+i\frac{q}{\epsilon}({\bf k.\hat{n}})\frac{\partial f}{\partial x^i}+\big(\dot{\eta}-\frac{1}{2}(\dot{h}+6\dot{\eta}({\bf k .\hat{n}})^2)\big) \frac{\partial f}{\partial q}=(\frac{\partial f}{\partial \tau})_c.
\end{equation}
The time evolution of a phase-space distribution function of the mother particles can be written according to the equation~\eqref{fmain2} as
\begin{eqnarray}\label{f}
&&(\frac{\partial f_{\rm DDM }}{\partial \tau})+i\frac{q_{\rm DDM }}{\epsilon_{\rm DDM }}({\bf k.\hat{n}}) \frac{\partial f_{\rm DDM }}{\partial x^i}+ \nonumber\\ 
&&(\dot{\eta}-\frac{1}{2}(\dot{h}+6\dot{\eta}({\bf k .\hat{n}})^2)\frac{\partial f_{\rm DDM }}{\partial q_{\rm DDM }}=(\frac{\partial f_{\rm DDM }}{\partial \tau})_c.
\end{eqnarray} 
To describe the decay process, we define the decay rate $\Gamma(q_D,q_{\rm DDM })f_{\rm DDM }$, which is the function
describing how many daughter particles with comoving momentum $q_D$ are created for a unit
time interval from the mother particles with momentum $q_{\rm DDM }$.
The collision term can be expressed as integration of $\Gamma(q_D,q_{\rm DDM })f_{\rm DDM }$ with $q_D$ as 
\begin{equation}
\left(\frac{\partial f_{\rm DDM }}{\partial \tau}\right)_c=\mp\int a\Gamma(q_D,q_{\rm DDM })f_{\rm DDM } d^3q_D,
\end{equation}
where we consider the minus sign as DM decay to other particles.
We can divide a distribution function $f_{\rm DDM }$ into the background (average), $f_{\rm DDM }^{(0)}$, and the perturbation $\Delta f_{\rm DDM }$ as follow
\begin{eqnarray}
&f_{\rm DDM }(q_{\rm DDM },{\bf{k,n}},\tau)=f_{\rm DDM }^{(0)}(q_{\rm DDM },\tau)+\nonumber\\
&\qquad\qquad\Delta f_{\rm DDM }(q_{\rm DDM },{\bf{k,n}},\tau).
\end{eqnarray}
By substituting the above relation into equation (\ref{f}), we obtain the Boltzmann equations for the mother particles at both the background and the perturbation levels as follows:\\	
\begin{equation}\label{fm}
\dot{f}_{\rm DDM }^{(0)}=-\frac{aM\Gamma}{E_{\rm DDM }}f_{\rm DDM }^{(0)},
\end{equation}
for the background where $E_{\rm DDM }=\sqrt{q^2_{\rm DDM }+M^2a^2}$. For the first order case we have
\begin{eqnarray}
&&\frac{\partial \Delta f_{\rm DDM }}{\partial \tau}+i \frac{q_{\rm DDM }}{ \epsilon_{\rm DDM }}({\bf k .\hat{n}})\Delta f_{\rm DDM }+q_{\rm DDM }\frac{\partial f_{\rm DDM }^{(0)}}{\partial q_{\rm DDM }}\nonumber\\
&&\left(\dot{\eta}-\frac{1}{2}(\dot{h}+6\dot{\eta}({\bf k .\hat{n}})^2)\right)=-\frac{aM\Gamma}{\epsilon_{\rm DDM }}\Delta f_{\rm DDM }.
\end{eqnarray}
For daughter particles, we have
\begin{equation}
\dot{f}_{\chi_i}^{(0)}=-\frac{a\Gamma N_{\rm DDM }(\tau)}{4\pi q^2}\delta(a-ap_{\rm max}),
\end{equation}
where $N_{\rm DDM } =(\Omega_{\rm DDM }^{\rm ini}\rho_{c,0}/M) \exp (-\Gamma t)$ is the mean comoving number density of the mother particles and $p_{\rm max}$ is the initial physical momentum of decay particles in the rest frame of
the mother particles, which is given by $p_{\rm max}=\frac{1}{2}[M^2-2(m_1^2+m_2^2)+\frac{(m_1^2+m_2^2)^2}{M^2}]^{1/2}$. We remind that collision terms are the same for both the daughter particles and we suppose that in  the conformal time $\tau_q$  the daughter particles with a comoving momentum $q=a(\tau_q)p_{\rm max}$ are produced~\citep{Aoyama:2014tga}.
An expansion of $\Delta f_{\rm DDM }$ in terms of the Legendre polynomials $P_l({\bf{\hat{k}.n}})$ with $l\geq0$ could define as 
\begin{equation}
\Delta f_{\rm DDM }=\Sigma_{l=0}^{+\infty}(-i)^l(2l+1)\Delta f_{\rm DDM }^{(l)}(q_{\rm DDM },\tau)P_l({\bf{\hat{k}.n}}).
\end{equation}
Since DDM  particles are non-relativistic, their zeroth-order phase-space distribution is the Maxwell-Boltzmann function so for the mother particles the higher-order multipole moments should vanish, i.e.,
\begin{equation}
\Delta f_{\rm DDM }^{(l)}=0\qquad for\quad l\geq 1,
\end{equation}
and for $l=0$, the monopole moment $\Delta f_{\rm DDM }^{(0)}$ obeys
the following equation,
\begin{equation}\label{fm2}
\frac{\partial \Delta f_{\rm DDM }}{\partial \tau}=-a\Gamma \Delta f_{\rm DDM }^{(0)}+\frac{1}{6}\dot{h}q_{\rm DDM }\frac{\partial f_{\rm DDM }^{(0)}}{\partial q_{\rm DDM }}.
\end{equation}
In order for the expansion of the universe to appear explicitly in the equations, we can trade $q$ for the physical momentum, $p$ so equations \eqref{fm} and \eqref{fm2} can be recast into evolution equations for the mean energy density $\bar{\rho}_{\rm DDM }$ and its perturbation $\bar{\rho}_{\rm DDM }\delta_{\rm DDM }$, which are defined as
\begin{equation}
\bar{\rho}_{\rm DDM }=\frac{1}{a^4}\int dq_{\rm DDM } 4\pi q_{\rm DDM }^2\epsilon_{\rm DDM }f_{\rm DDM }^{(0)},
\end{equation}
\begin{equation}
\bar{\rho}_{\rm DDM }\delta_{\rm DDM }=\frac{1}{a^4}\int dq_{\rm DDM } 4\pi q_{\rm DDM }^2\epsilon_{\rm DDM }\Delta f_{\rm DDM }^{(0)}.
\end{equation}
We can obtain by integrating equations \eqref{fm} and \eqref{fm2} and multiply by $4\pi q_{\rm DDM }^2\epsilon_{\rm DDM }/a^4$:
\begin{equation}
\frac{d}{d\tau}\bar{\rho}_{\rm DDM }+3{\cal H}\bar{\rho}_{\rm DDM }=-a\Gamma \bar{\rho}_{\rm DDM },\\
\end{equation}
\begin{equation}
\frac{d}{d\tau}[\bar{\rho}_{\rm DDM }\delta_{\rm DDM }]+3{\cal H}\bar{\rho}_{\rm DDM }\delta_{\rm DDM }=-\frac{\dot{h}}{2}\bar{\rho}_{\rm DDM }-a\Gamma \bar{\rho}_{\rm DDM },\\
\end{equation}
and we combine these two equations lead to
\begin{eqnarray}
\dot{\delta}_{\rm DDM }&=-\frac{\dot{h}}{2},\\
\dot{\theta}_{\rm DDM }&=-{\cal{H} }\theta_{\rm DDM },
\end{eqnarray}
that $\delta$ is the dimensionless perturbation $\theta_{\rm DDM }=\partial_i v_{\rm DDM }^i=0$ in the synchronous gauge.
We note according to the above equation, mother particles are the same as that for CDM without decay and for them $\bar{\rho}_{\rm DDM }\sim a^{-3}$, since the mother particles are non-relativistic~\citep{Poulin:2016nat}.\\
We can  express previous equations in the Newtonian gauge by using relations~\citep{Lesgourgues:2011rh}:
\begin{eqnarray}
&&\delta^{(N)}=\delta^{(S)}+\frac{\rho'}{\rho}\alpha\\
&&\theta^{(N)}=\theta^{(S)}+k^2\alpha,
\end{eqnarray}
where $\alpha=(\dot{h}+6\dot{\eta})/2k^2$,as 
\begin{eqnarray}
&&\delta^{(N)}=\delta^{(S)}+(3{\cal{H}}+a\Gamma)\alpha\\
&&\theta^{(N)}=\theta^{(S)}+k^2\alpha.
\end{eqnarray}
We introduce the gauge invariant variables given in Table~\ref{tab-sn} and can write these equations as
\begin{eqnarray}
&&\dot{\delta}_{\rm DDM }=-\theta_{\rm DDM }-m_{\rm cont}-a\Gamma m_{\psi}, \nonumber\\
&&\dot{\theta}_{\rm DDM }=-\frac{\dot{a}}{a}\theta_{\rm DDM }+k^2 m_{\psi}, \nonumber
\end{eqnarray}
\begin{table}
	\centering
	\caption{Metric source terms for scalar perturbations in synchronous and Newtonian gauge.}
	\begin{tabular}{|c | c |c|}
		\hline 
		&Synchronous& Newtonian\\
		\hline
		$m_{\rm cont}$&$\dot{h}/2$&$-3\dot{\phi}$\\
		$m_{\psi}$&$0$&$\psi$\\
		$m_{\rm shear}$&$(\dot{h}+6\dot{\eta})/2$&$0$\\
		\hline 
	\end{tabular}\label{tab-sn}
\end{table}
Similarly, the Boltzmann equations of daughter particles are
\begin{equation}\label{fD}
\frac{\partial f_{\rm j}}{\partial \tau}+\frac{dx^i}{d\tau} \frac{\partial f_{\rm j}}{\partial x^i}+\frac{dq_{\rm }}{d\tau} \frac{\partial f_{\rm j}}{\partial q_{\rm }}+\frac{dn^i}{d\tau} \frac{\partial f_{\rm DDM }}{\partial n^i}=(\frac{\partial f_{\rm j}}{\partial \tau})_c,
\end{equation}
where j is the particle index.
 The collision term can be written as
\begin{eqnarray}\label{col-d}
&&	\left(\frac{\partial f_1}{\partial \tau}\right)_c=\frac{a (M^2+m_1^2-m_2^2)}{2M^2}\Gamma f_{\rm DDM }^{(0)},\\
&&	\left(\frac{\partial f_2}{\partial \tau}\right)_c=\frac{a (M^2+m_2^2-m_1^2)}{2M^2}\Gamma f_{\rm DDM }^{(0)}.\label{col-d2}
\end{eqnarray}
The factors $\frac{(M^2+m_1^2-m_2^2)}{2M^2}$ and $\frac{(M^2+m_2^2-m_1^2)}{2M^2}$ that appear in the above collision terms can be easily understood. Consider a two-body decay in the rest frame
of the DDM  particle with mass M and the masses corresponding to the daughter particles are indicated by $m_1$ , and $m_2$ . The energies of the daughter particles in the rest frame of DDM  are $E_1 = (M^2+m_1^2-m_2^2 )/2M^2$
and $E_2 = (M^2+m_2^2-m_1^2 )/2M^2$. So these factors represent the ratios of energy that have been deposited into different daughter particle species.
By defining the mass of each daughter particle to the DDM  particle, $\gamma_i=\frac{m_i^2}{M^2}$ and $\epsilon=\frac{1}{2}(1+\gamma_1^2-\gamma_2^2)$can be rewritten the equations \eqref{col-d} and \eqref{col-d2} as
\begin{eqnarray}\label{eq-coll}
&&	\left(\frac{\partial f_1}{\partial \tau}\right)_c=a\epsilon\Gamma f_{\rm DDM }^{(0)},\\
&&	\left(\frac{\partial f_2}{\partial \tau}\right)_c=a (1-\epsilon) \Gamma f_{\rm DDM }^{(0)}.
\end{eqnarray} 
In all classes of DDM  models, there are two parameters: decay rate $\Gamma$ (or decay lifetime $\Gamma^{-1}$); and the fraction of DDM  rest mass energy converted into daughter mass energies $\epsilon$ which $0\leq \epsilon \leq 1$ and $0\leq \epsilon \leq 1/2$ are for the first two scenarios and two last models with massless particles(DR), respectively.\\	
The daughter particles must be treated differently to account for their finite mass and non-trivial velocity kicks. 
If we restrict attention  to the first case (i) introduced in section~\ref{sec:int} in which the mass
difference between the DDM  and daughter particles are small, so $f = 1-m_i/M \leq 1$, the daughter particle will receive an extremely non-relativistic kick velocity. As we should expect, they behave similarly to CDM.
In this limit, the daughter perturbations evolve as for a standard non-relativistic dark matter species. So in this part similar to mother particles, DDM, we again divide the distribution function of massive daughter particles, $CDM_i$ into the background $f_{\rm i}^{(0)}$ and the perturbation $\Delta f_{\rm i}$  as
\begin{equation}
f_{\rm i}(q,{\bf{k,n}},\tau)=f_{\rm i}^{(0)}(q,\tau)+\Delta f_{\rm i}(q,{\bf{k,n}},\tau);
\end{equation}
and we write $\Delta f_{\rm i}^{(l)}$ as the l-th multipole moment of $\Delta f_j$ as
\begin{equation}\label{pl}
\Delta f_{\rm i}=\Sigma_{l=0}^{+\infty}(-i)^l(2l+1)\Delta f_{\rm i}^{(l)}(q,\tau)P_l({\bf{\hat{k}.n}}).
\end{equation}
and we can obtain
\begin{equation}
\dot{\delta}_{\rm CDM1}=-(\theta_{\rm CDM1}+m_{\rm cont})+a \Gamma \epsilon\frac{\rho_{\rm DDM }}{\rho_{\rm CDM1}}(\delta_{\rm DDM }-\delta_{\rm CDM1}+m_\psi),\nonumber
\end{equation}
\begin{equation}
\dot{\theta}_{\rm CDM1}=-\frac{\dot{a}}{a}\theta_{\rm CDM1}+k^2m_\psi-a \Gamma \epsilon\frac{\rho_{\rm DDM }}{\rho_{\rm CDM1}}\theta_{\rm CDM1}.
\end{equation}
and for another daughter particles as
\begin{equation}
\dot{\delta}_{\rm CDM2}=-(\theta_{\rm CDM2}+m_{\rm cont})+a \Gamma (1-\epsilon)\frac{\rho_{\rm DDM }}{\rho_{\rm CDM2}}(\delta_{\rm DDM }-\delta_{\rm CDM2}+m_\psi),
\end{equation}
\begin{equation}
\dot{\theta}_{\rm CDM2}=-\frac{\dot{a}}{a}\theta_{\rm CDM2}+k^2m_\psi-a \Gamma (1-\epsilon)\frac{\rho_{\rm DDM }}{\rho_{\rm CDM2}}\theta_{\rm CDM2}.
\end{equation}
We can obtain their continuity equation as 
\begin{eqnarray}\label{rhosdm}
&&\dot{\rho}_{\rm CDM1}+3{\cal{H}}\rho_{\rm CDM1}=\Gamma a\epsilon\rho_{\rm DDM },\nonumber\\
&&\dot{\rho}_{\rm CDM2}+3{\cal{H}}\rho_{\rm CDM2}=\Gamma a (1-\epsilon)\rho_{\rm DDM }.
\end{eqnarray}
In the second case(WDM-CDM), we follow the procedure used in \cite{Lesgourgues:2011rh} and \cite{Abellan:2021bpx} as they introduced a novel approximation
scheme to compute dynamics of the WDM linear perturbations accurately and quickly by considering the WDM species as a viscous fluid on sub-Hubble scales. By performing similar calculations for the warm daughter particles, the following equations are obtained:
\begin{eqnarray}\label{eq:wdm}
&&\dot{\delta}_{\rm WDM}=-3{\cal{H}}(c_s^2-w)\delta_{\rm WDM}-(1+w)(\theta_{\rm WDM}+m_{\rm cont})\nonumber\\
&&+a \Gamma (1-\epsilon)\frac{\rho_{\rm DDM }}{\rho_{\rm WDM}}(\delta_{\rm DDM }-\delta_{\rm WDM}+m_\psi),\\
&&\dot{\theta}_{\rm WDM}=-{\cal{H}}(1-3c_g^2)\theta_{\rm WDM}+\frac{c_s^2}{1+w}k^2m_\psi-k^2\sigma_{\rm WDM},\nonumber\\
&&-a \Gamma (1-\epsilon)\frac{1+c_g^2}{1+w}\frac{\rho_{\rm DDM }}{\rho_{\rm WDM}}\theta_{\rm WDM},
\end{eqnarray}
where w is the dynamical equation of state of the massive warm daughter particle that could be written in terms of the mean speed of a massive daughter particle.
Here, $c_g^2\equiv\frac{\dot{p}}{\dot{\rho}}$ is the WDM adiabatic sound speed and $c_s^2\equiv\frac{\delta p}{\delta \rho}$ is WDM sound speed  in the synchronous gauge. The adiabatic sound speed be expressed in ~\citep{Abellan:2021bpx} as
\begin{eqnarray}
&&c_g^2=w\left(5-\frac{{\cal{P}}_{\rm WDM}}{p_{\rm WDM}}-\frac{\rho_{\rm DDM }}{\rho_{\rm WDM}}\frac{a\Gamma}{3w{\cal{H}}}\frac{\epsilon^2}{1-\epsilon}\right)\nonumber\\
&&\times \left[3(1+w)\frac{\rho_{\rm DDM }}{\rho_{\rm WDM}}\frac{a\Gamma}{{\cal{H}}}(1-\epsilon)\right]^{-1}
\end{eqnarray}
where the quantity ${\cal{P}}_{\rm WDM}$ (called the pseudo-pressure inside \texttt{CLASS}) is a higher moment pressure that is reduced to the
standard pressure in the relativistic limit. On the other hand, since there is no dynamic equation for pressure perturbation, obtaining an analytical expression for $c^2_s$ is more complicated so in Ref. ~\citep{Abellan:2021bpx} and ~\citep{Lesgourgues:2011rh}, it is supposed that $c^2_s$ is scale-independent and approximately equal to $c^2_g$. But calculations using the full Boltzmann hierarchy show that reveals that $c^2_s$ represents a specific k-dependence and cannot be obtained with a background quantity such as $c^2_g$ and it increases slightly on the scales k. 
For the synchronous sound speed, we follow the prescription in
\begin{equation}
c^2_s(k)=c^2\left[1+\frac{1-2\epsilon}{5}\sqrt{\frac{k}{k_{fs}}}\right],
\end{equation}
where $k_{fs}=\sqrt{\frac{3}{2}{\cal{H}}/c_g}$ is the free-streaming length of the WDM particles.
Here, we use the same method as the references mentioned above and we implemented the corresponding equations similar to the available code\footnote{\url{https://github.com/PoulinV/class_majoron}}. 
In equation~\ref{eq:wdm}, $\sigma_{\rm WDM}$ represents the WDM shear perturbations and are negligible for the warm particles for simplicity same CDM. 
 The continuity equation for WDM particles obtained as
\begin{equation}
\dot{\rho}_{\rm WDM}+3{\cal{H}}(1+w)\rho_{\rm WDM}=\Gamma a (1-\epsilon)\rho_{\rm DDM }.
\end{equation}
In the third case (CDM-DR), one of the daughter particles resulting from decay is dark radiation so the perturbations for the massless relativistic daughter particles need to write the full Boltzmann hierarchy. In other words, they may be treated in a form analogous to that of massless neutrinos. We integrated the phase-space distribution function and expand them over Legendre polynomials $P_l$ in the following way
\begin{equation}
F_{\rm DR}\equiv\frac{\int dq q^3 f_{\rm DR}^{(0)}\Psi_{\rm DR}}{\int dq q^3 f_{\rm DR}^{(0)}}r_{\rm DR},
\end{equation}
where $ \Psi_{\rm DR} $  is defined at the level of the perturbed phase-space distribution:
\begin{equation}
f_{\rm DR}(x,q,\tau)=f_{\rm DR}^{(0)}(q,\tau)(1+\Psi_{\rm DR}(x,q,\tau)),
\end{equation}
and $r_{\rm DR}$ defined as
\begin{equation}
r_{\rm DR}\equiv\frac{\rho_{\rm DR} a^4}{\rho_{\rm cr,0}},
\end{equation}
here to make $r_{\rm DR}$  dimensionless, the critical energy density today, $\rho_{\rm cr,0}$ has been applied~\citep{Audren:2014bca}. 
Evaluating the Boltzmann equation for Legendre polynomial expansion for mass-less particles  yields the evolution of the multipole coefficients in the conventional
notation,
\begin{eqnarray}
&&\dot{F}_{\rm DR,0}=-kF_{\rm DR,1}-\frac{4}{3}r_{\rm DR}m_{\rm cont}+\dot{r}_{\rm DR}(\delta_{\rm DDM }+m_{\psi}),\nonumber\\
&&\dot{F}_{\rm DR,1}=\frac{k}{3}F_{\rm DR,0}-\frac{2k}{3}F_{\rm DR,2}+\frac{4k}{3}r_{\rm DR} m_{\psi}+\frac{\dot{r}_{\rm DR}}{k}\theta_{\rm DDM },\nonumber\\
&&\dot{F}_{\rm DR,2}=\frac{2k}{5}F_{L,1}-\frac{3k}{5}F_{\rm DR,3}+\frac{8}{15}r_Lm_{\rm shear},\nonumber\\
&&\dot{F}_{{\rm DR},l}=\frac{k}{2l+1}(lF_{{\rm DR},l-1}-(l+1)F_{{\rm DR},l+1})\nonumber\\
&&\qquad\quad-a\Gamma (1-\epsilon)\frac{\rho_{\rm DDM }}{\rho_{\rm DR}}F_{{\rm DR},l}\qquad l\geq 3,
\end{eqnarray}
that the derivative of $r_{\rm DR}$ is given by
\begin{equation}
\dot{r}_{\rm DR}=a (1-\epsilon)\Gamma \frac{\rho_{\rm DDM }}{\rho_{\rm DR}}r_{\rm DR}.
\end{equation}
The relation of the multipole moments to the standard variables $ \delta $ and $ theta $ are given by:
\begin{equation}
F_{\rm DR,0}=r_{\rm DR}\delta_{\rm DR},  F_{\rm DR,1}=\frac{4}{3k}r_{\rm DR} \theta_{\rm DR}, F_{\rm DR,2}=2\sigma r_{\rm DR}.
\end{equation}
The evolution of the multipole coefficients  obtain as
\begin{eqnarray}
\dot{\delta}_{\rm DR}&=&-\frac{4}{3}(\theta_{\rm DR}+m_{\rm cont})+\nonumber\\
&&a\Gamma (1-\epsilon)\frac{\rho_{\rm DDM }}{\rho_{\rm DR}}(\delta_{\rm DDM }-\delta_{\rm DR}+m_\psi),\nonumber\\
\dot{\theta}_{\rm DR}&=&k^2(\frac{\delta_{\rm DR}}{4}-\sigma_{\rm DR}+m_\psi)\nonumber\\
&&-a\Gamma(1-\epsilon)\frac{3\rho_{\rm DDM }}{4\rho_{\rm DR}}(\frac{4}{3}\theta_{\rm DR}-\theta_{\rm DDM }),\nonumber\\
\dot{\sigma}_{\rm DR}&=&\frac{4}{15}(\theta_{\rm DR}+m_{\rm shear}-\frac{9}{8}kF_{\rm DR,3})-\nonumber\\
&&a\Gamma(1-\epsilon)\frac{\rho_{\rm DDM }}{\rho_{\rm DR}}\sigma_{\rm DR},
\end{eqnarray}	
and the continuity equation of dark radiation obtain as 
\begin{equation}\label{rhol}
\dot{\rho}_{\rm DR}+4{\cal{H}}
\rho_{\rm DR}=\Gamma a(1-\epsilon)\rho_{\rm DDM }.
\end{equation}
In the last case(WDM-DR), we consider a massive cold parent particle decaying
to one massless (DR) and one massive warm daughter particle. Such models could arise beyond the Standard Model that includes Super WIMPs or excited dark fermions with magnetic dipole transitions. Here, $\epsilon$  is the fraction of the rest mass energy of the parent particle that is transferred to the massless daughter.\\
In the end of this section, we point in all cases and for a flat universe through the Friedmann equation we can write the Hubble parameter as 
\begin{equation}
{\cal{H}}^2(a)=\frac{8\pi G a^2}{3}\left(\rho_{\rm DDM }(a)+\rho_i(a)+\rho_\gamma a^{-4}+\rho_b a^{-3}+\rho_\Lambda\right),
\end{equation}
where $\rho_\gamma,\rho_b, \rho_\Lambda$ denote the mean densities
of photons, baryons and dark energy, respectively and $\rho_i$ assign to particles caused by decay means $\rho_{\rm CDM},\rho_{\rm WDM}$, and $\rho_{\rm DR}$.
\begin{table}
	\centering
	\caption{The best values of the free parameters obtained
		by considering Planck TT,TE,EE+lowE+lensing~\citep{Aghanim:2018eyx} for different scenarios decaying into two particles. It is obvious from these values that the tensions are not solved.} 

		CDM-CDM
		\begin{tabular}{|l|c|c|}
				\hline 
		Param & best-fit & mean$\pm\sigma$  \\ \hline 
$\Omega_\mathrm{B}$ &$0.04811$ & $0.04856_{-0.00071}^{+0.00066}$ \\ 
$\Omega_\mathrm{DM}$ &$0.2509$ & $0.2558_{-0.0077}^{+0.0073}$ \\ 
$100\theta_{MC}$ &$1.042$ & $1.042_{-0.00032}^{+0.00032}$ \\ 
$\ln{10}^{10}A_s$ &$3.052$ & $3.057_{-0.019}^{+0.016}$  \\ 
$n_s$ &$0.9736$ & $0.9734_{-0.0047}^{+0.0047}$ \\ 
$\tau_{reio}$ &$0.05774$ & $0.05946_{-0.0092}^{+0.0079}$ \\ 
$H_0$ &$68.46$ & $68.08_{-0.64}^{+0.61}$  \\ 
$\sigma_8$ &$0.8073$ & $0.8125_{-0.0078}^{+0.0075}$  \\ 
$\Gamma$ &$7.751$ & $ <154$\\ 
$\epsilon$ &$0.5936$ & $0.5043_{-0.21}^{+0.23}$ \\

		\hline 
	\end{tabular} \label{tabcdm}
	CDM-WDM	
	\begin{tabular}{|l|c|c|}
		
		\hline 
		Param & best-fit & mean$\pm\sigma$  \\ \hline 
$\Omega_\mathrm{B}$ &$0.04828$ & $0.04765_{-0.00076}^{+0.0015}$  \\ 
$\Omega_\mathrm{DM}$&$0.253$ & $0.2449_{-0.0098}^{+0.016}$  \\ 
$100\theta_{MC}$&$1.042$ & $1.042_{-0.00033}^{+0.00034}$ \\ 
$\ln{10}^{10}A_s$ &$3.057$ & $3.064_{-0.019}^{+0.016}$  \\ 
$n_s$ &$0.9764$ & $0.9718_{-0.005}^{+0.0044}$\\ 
$\tau_{reio}$ &$0.05925$ & $0.06233_{-0.0093}^{+0.0087}$  \\ 
$H_0$ &$68.23$ & $68.69_{-1.2}^{+0.67}$ \\ 
$\sigma_8$  &$0.8113$ & $0.8288_{-0.02}^{+0.0091}$ \\ 
$\Gamma$ &$34.34$ & $< 76.9$ \\
$\epsilon$ &$0.0998$ & $0.0501_{-0.014}^{+0.051}$  \\
$m_{\rm CDM}$ &$0.3677$ & $0.5101_{-0.29}^{+0.26}$ \\
		\hline 
	\end{tabular}  \label{tabwdm}
CDM-DR
		\begin{tabular}{|l|c|c|} 
			
		\hline 
		Param & best-fit & mean$\pm\sigma$  \\ \hline 
$\Omega_\mathrm{B}$ &$0.0485$ & $0.0483_{-0.0008}^{+0.00077}$\\ 
$\Omega_\mathrm{DM}$ &$0.2555$ & $0.2522_{-0.0088}^{+0.0087}$\\ 
$100\theta_{MC}$ &$1.042$ & $1.042_{-0.00033}^{+0.0003}$ \\ 
$\ln{10}^{10}A_s$ &$3.046$ & $3.054_{-0.018}^{+0.019}$  \\ 
$n_s$ &$0.9695$ & $0.974_{-0.0049}^{+0.005}$ \\ 
$\tau_{reio}$ &$0.05423$ & $0.05818_{-0.0096}^{+0.0085}$ \\ 
$H_0$ &$68.06$ & $68.25_{-0.72}^{+0.68}$ \\ 
$\sigma_8$ &$0.8067$ & $0.8092_{-0.0085}^{+0.0084}$\\
$\Gamma$ &$9.898$ & $<31.9$  \\ 
$\epsilon$ &$0.9869$ & $0.7346_{-0.052}^{+0.27}$ \\ 
		\hline 
	\end{tabular}
	WDM-DR
	\begin{tabular}{|l|c|c|} 
		\hline 
		Param & best-fit & mean$\pm\sigma$  \\ \hline 
$\Omega_\mathrm{B}$ &$0.04881$ & $0.04829_{-0.00056}^{+0.00078}$ \\ 
$\Omega_\mathrm{DM}$ &$0.2572$ & $0.249_{-0.0064}^{+0.011}$ \\
$100\theta_{MC}$ &$1.042$ & $1.042_{-0.00027}^{+0.00029}$  \\ 
$\ln{10}^{10}A_s$ &$3.062$ & $3.06_{-0.017}^{+0.017}$ \\ 
$n_s$ &$0.9743$ & $0.9721_{-0.004}^{+0.004}$  \\ 
$\tau_{reio}$ &$0.06098$ & $0.06011_{-0.0088}^{+0.0082}$  \\ 
$H_0$ &$67.92$ & $68.21_{-0.69}^{+0.49}$   \\ 
$\sigma_8$ &$0.8176$ & $0.8194_{-0.01}^{+0.0073}$  \\ 
$\Gamma$ &$0.4781$ & $<2.78$ \\  
$\epsilon$ &$0.1206$ & $<0.43$  \\
		\hline 
	\end{tabular} \label{tabwdmdr} 
\end{table}
\begin{table}
	\centering
	\caption{The mean (best-fit) $\pm1\sigma$ errors of the cosmological parameters for $\Lambda$CDM model from Planck TT,TE,EE+lowE+lensing.}
	
	\begin{tabular}{|l|c|c||} 
		\hline 
		Param & best-fit & mean$\pm\sigma$ \\ \hline 
$\Omega_\mathrm{B}$ &$0.04773$ & $0.04864_{-0.00057}^{+0.00067}$ \\
$\Omega_\mathrm{DM}$ &$0.2468$ & $0.2567_{-0.0068}^{+0.0071}$  \\ 
$100\theta_{MC}$ &$1.042$ & $1.042_{-0.00032}^{+0.00029}$   \\ 
$\ln{10}^{10}A_s$ &$3.069$ & $3.058_{-0.017}^{+0.014}$ \\ 
$n_s$  &$0.9773$ & $0.9726_{-0.0045}^{+0.0039}$\\ 
$\tau_{reio}$ &$0.06908$ & $0.05948_{-0.0082}^{+0.0074}$ \\ 
$H_0$ &$68.77$ & $68_{-0.6}^{+0.55}$   \\ 
$\sigma_8$ &$0.812$ & $0.8133_{-0.0062}^{+0.0064}$ \\
		
		\hline 
	\end{tabular} \label{tablcdm}
\end{table}
\begin{table*}
	\caption{The result of MCMC analysis for all reviewed models in this study.} 
	\begin{tabular}{l|c|c|c|c|c|}
		\hline
		Model&$\Lambda$CDM&CDM-CDM&CDM-WDM&CDM-DR&WDM-DR\\
		\hline
		$\chi^2_{\rm min}$&$2991$&$2991$&$2991$&$2992$&$2990$\\
		$-\ln{\cal{L}}_\mathrm{min}$&$1495.37$&$1495.38$&$1495.32$&$1496$&$1494.85$\\
		\hline
	\end{tabular}
\end{table*}
\section{Observable effects}\label{sec-data}
\begin{figure}
	\begin{center}
\includegraphics[width=8cm]{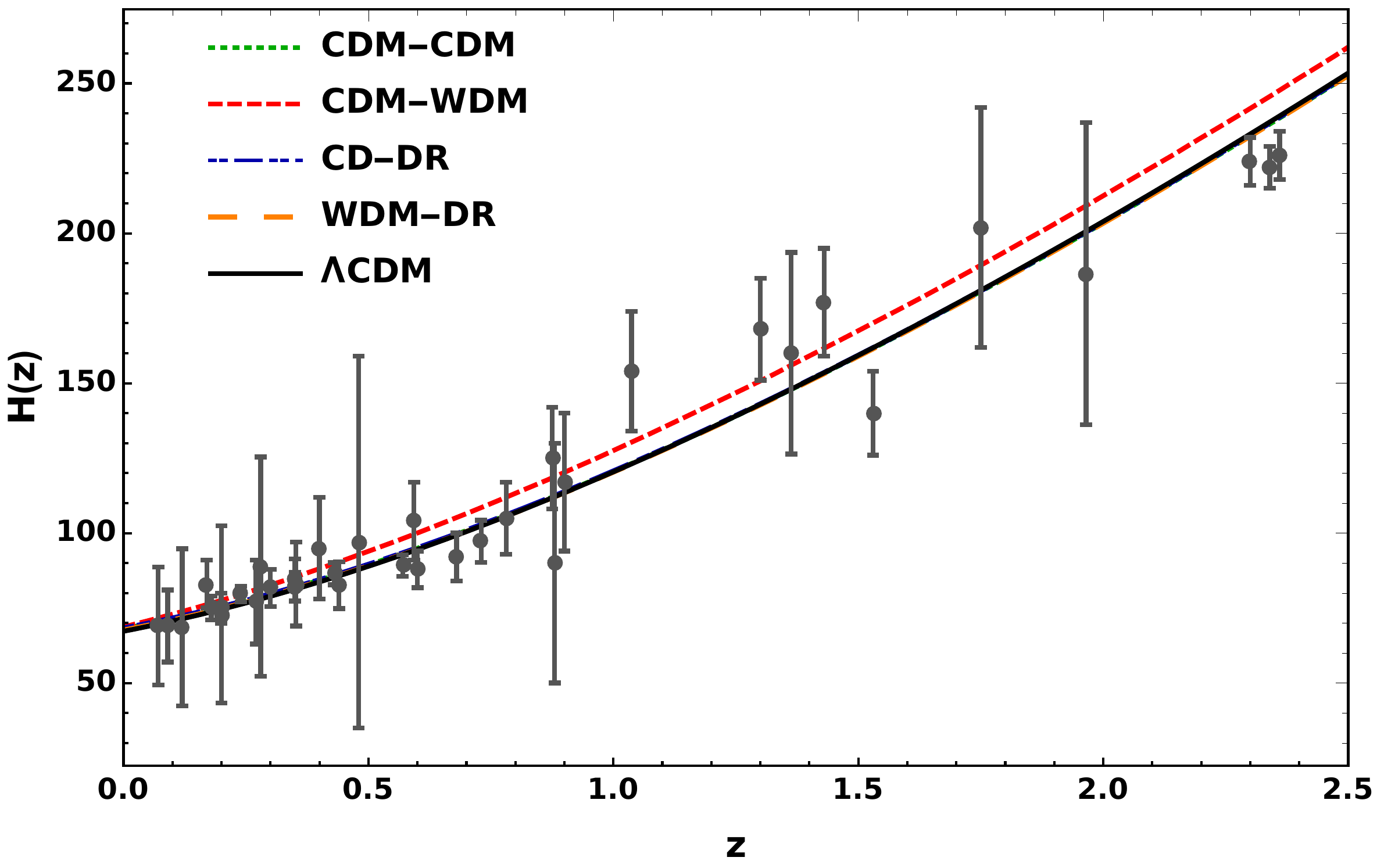}
		\caption{Theoretical predicted the Hubble parameter as a function of redshift using the best fit values
			of cosmological parameters in Tables \ref{tabwdmdr} and \ref{tablcdm}  for the proposed decaying dark matter scenarios and $\Lambda$CDM model compared to the observational data from cosmic chronometers~\citep{Marra:2017pst}.
			} 
		\label{figH}
	\end{center}
\end{figure}

\begin{figure*}
	\begin{center}
		\includegraphics[width=18cm]{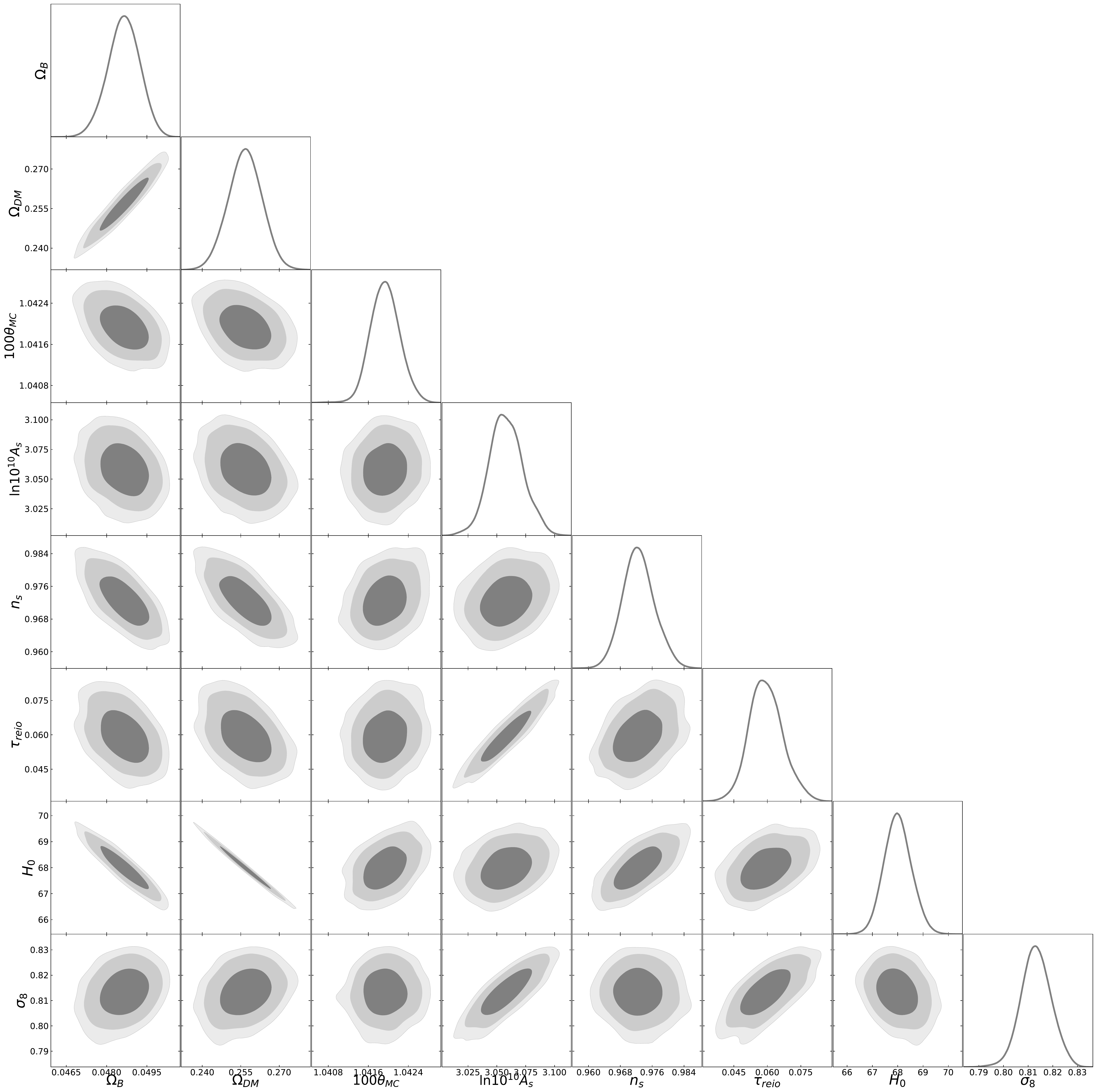}
		
		\caption{Reconstructed 2D posterior distribution in the $\Lambda$CDM model.} 
		\label{figlcdm}
	\end{center}
\end{figure*}

\begin{figure*}
	\begin{center}
		\includegraphics[width=18cm]{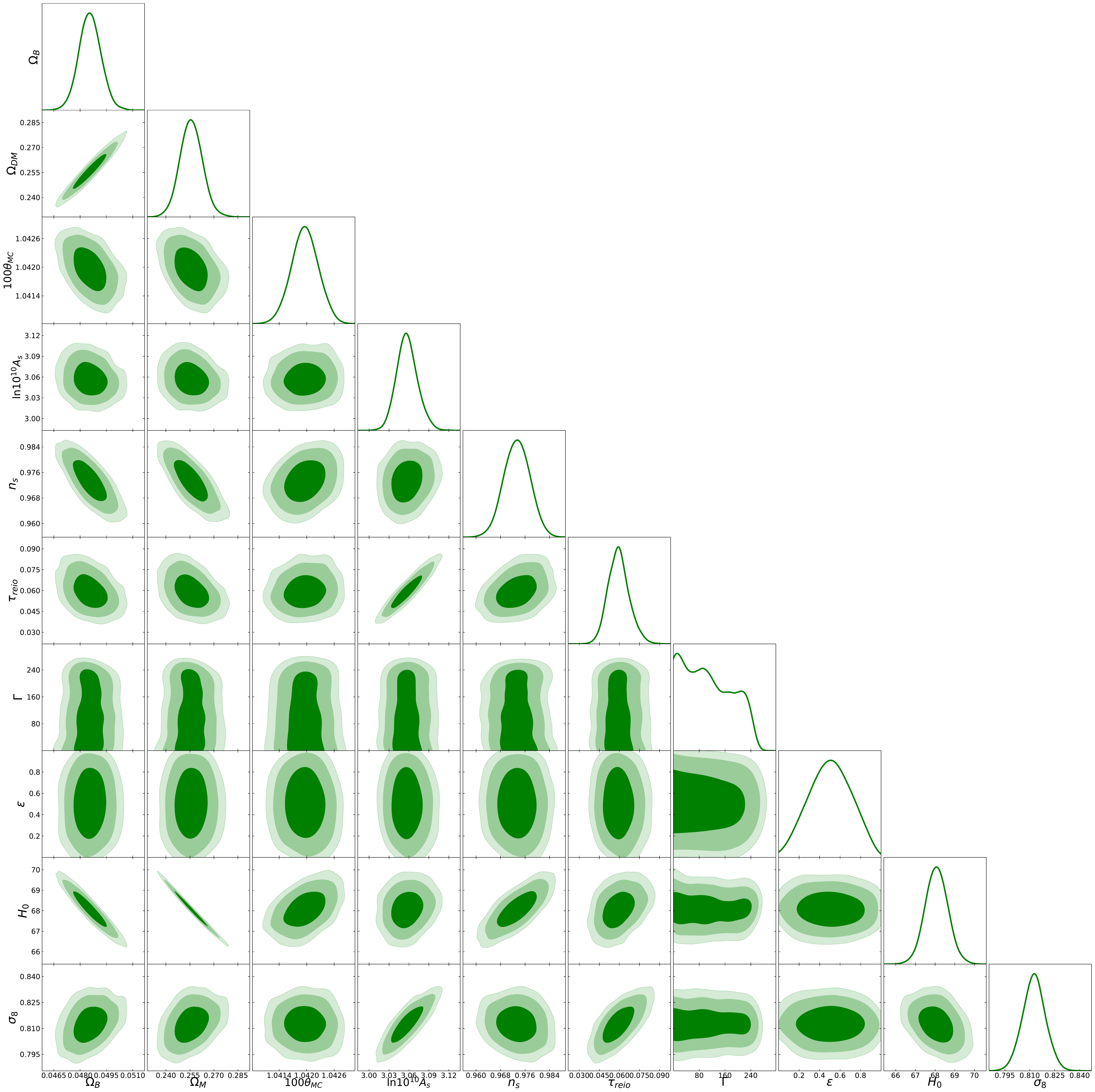}
		
		\caption{Reconstructed 2D posterior distribution in the CDM-CDM model.} 
		\label{figcdm}
	\end{center}
\end{figure*}

\begin{figure*}
	\begin{center}
		\includegraphics[width=18cm]{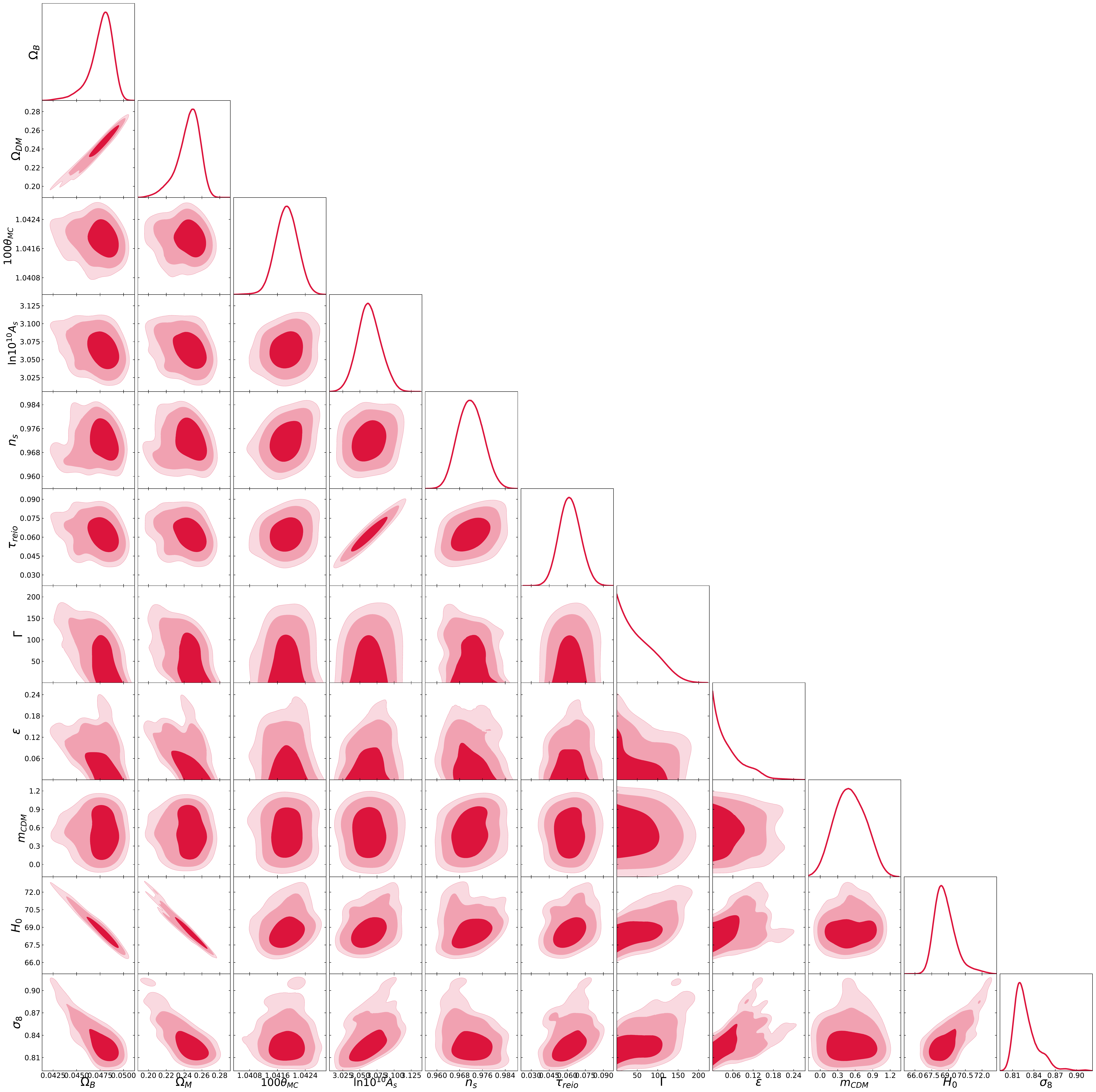}
		
		\caption{Reconstructed 2D posterior distribution in the CDM-WDM model.} 
		\label{figwdm}
	\end{center}
\end{figure*}

\begin{figure*}
	\begin{center}
		\includegraphics[width=18cm]{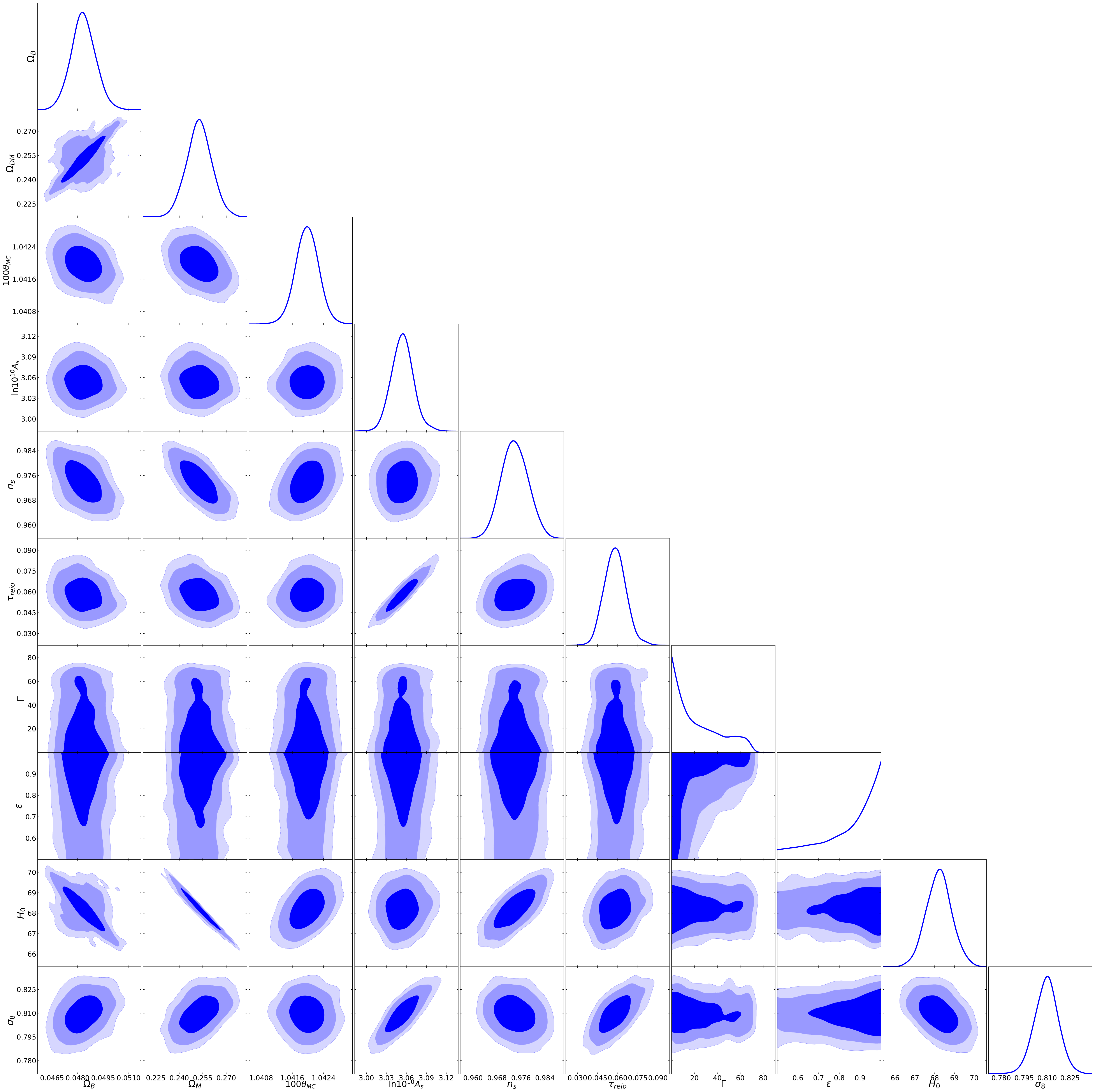}
		
		\caption{Reconstructed 2D posterior distribution in the CDM-DR model.} 
		\label{figcdmdr}
	\end{center}
\end{figure*}

\begin{figure*}
	\begin{center}
		\includegraphics[width=18cm]{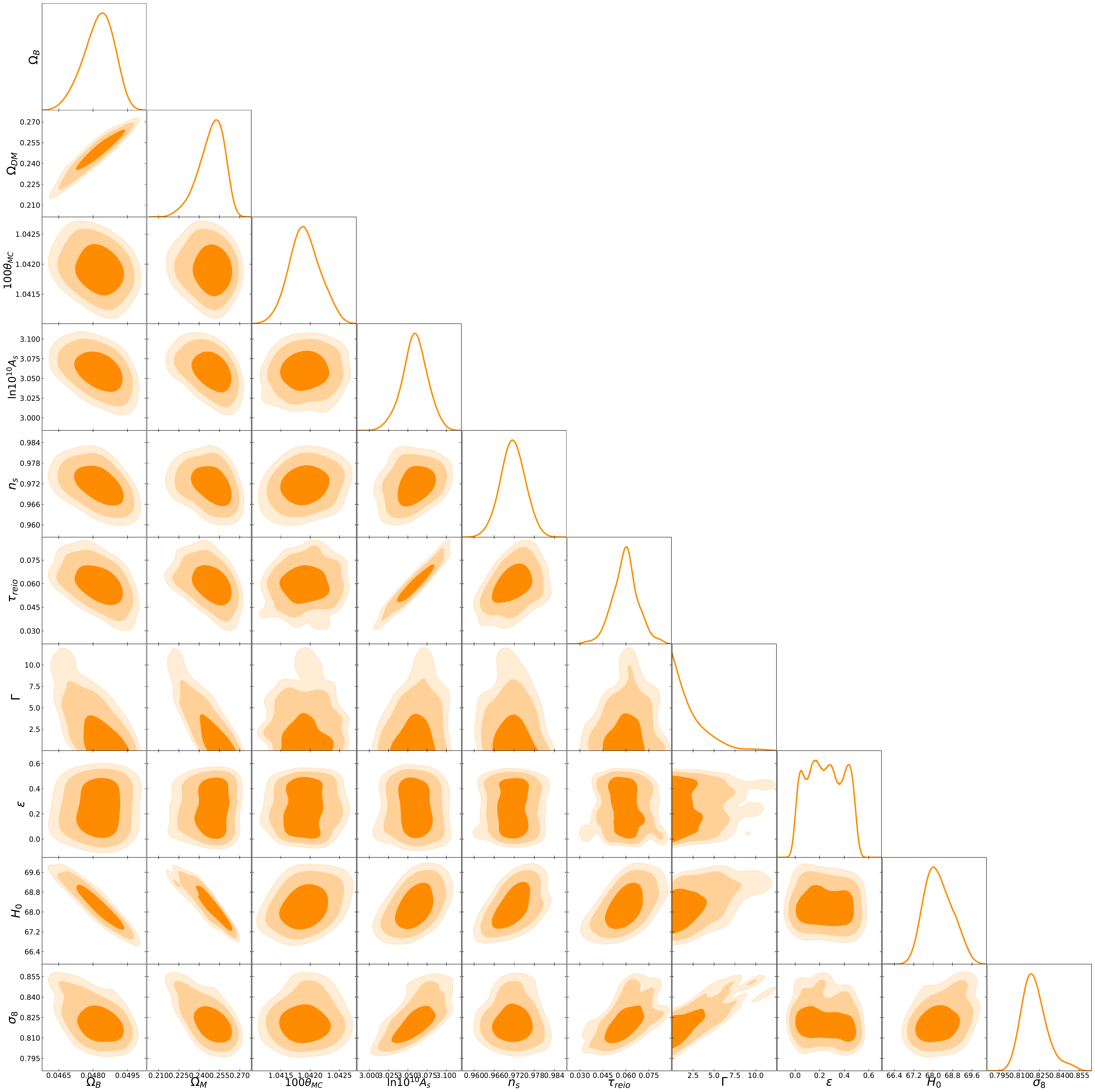}
		
		\caption{Reconstructed 2D posterior distribution in the WDM-DR model.} 
		\label{figwdmdr}
	\end{center}
\end{figure*}

The whole free parameter space in all mentioned models in the previous section is specified with the  cosmological parameters: $(\Omega_b,\Omega_{DM},10\theta_{\rm MC},\ln 10^{10}A_s,n_s,\tau_{\rm reio}, \Gamma)$, which are in order the density parameters of baryons and mother particles, angular parameter, amplitude of the primordial curvature perturbation at k = 0.002 Mpc$^{-1}$, the spectral index and, the optical depth of reionization. For the second scenario, CDM-WDM according to the CLASS package we modified, the mass of both decay particles is required so the
mass of the cold daughter particle, $m_{\rm CDM}$ is considered as a free parameter and the mass of the
other particle obtain through $\epsilon$. According to that, the mother particles are assumed to be cold, so we
choose to normalize it to 1.
We consider decay constant of $\Gamma$ is $\rm kms^{-1}Mpc^{-1}$, same unit as $H_0$ in \texttt{CLASS}. We recall that $ 1 km s^{-1} Mpc^{-1} = 1.02\times10^{-3} \rm Gyr^{-1}$ for translation with other works making use of $\rm Gyr ^{-1}$.
In this part, we will constrain our models with CMB observations so implement the DDM  equations in the publicly available numerical code \texttt{CLASS} and we use the shooting method to compute the present-day dark matter density described in~\cite{Audren:2014bca}. We use the publicly available Markov chain Monte Carlo code \texttt{MontePython-v3}
interfaced with our modified version of \texttt{CLASS} to explore free parameters with the flat priors for the decay rate and the mass-ratio of the mother and daughter particles. 
We could compare alternative models with the data simply by checking to see that they reproduce the peak positions and amplitudes~\citep{WMAP:2003tof}. So in this study, when studying decaying dark matter scenarios, we constrain the free parameters by using Planck TT,TE,EE+lowE+lensing~\citep{Aghanim:2018eyx}. We assume chains to be converged with the Gelman-Rubin convergence criterion
$R-1 < 0.05$~\citep{Gelman:1992zz}. We reported the best fit values of free parameters for the different scenarios of DDM   and
$\Lambda$CDM model from MCMC analysis in Tables~\ref{tabwdmdr} and ~\ref{tablcdm}. We also show 1$\sigma$ and 2$\sigma$ posterior distributions for the parameters in Figures~\ref{figlcdm} to~\ref{figwdmdr} by using \texttt{Getdist}~\citep{Lesgourgues:2011rh}.\\
To select the model that is most consistent with the observational data, we need a method that can numerically determine whether the fit is good. The least squares method ($\chi^2_{\rm min}$) is the simplest method commonly used in cosmology, especially  for comparing different models with the same number of parameters, it is sufficient and very popular. In this case, the model with a smaller $\chi^2_{\rm min}$ means it had a
better fit with the data. In this study, except for the second scenario(CDM-WDM), all models intended for dark matter decay have two additional parameters than the standard model and they all fit equally to cosmological data and the standard $\Lambda$CDM model.
\begin{figure}
	\begin{center}
		\includegraphics[width=8cm]{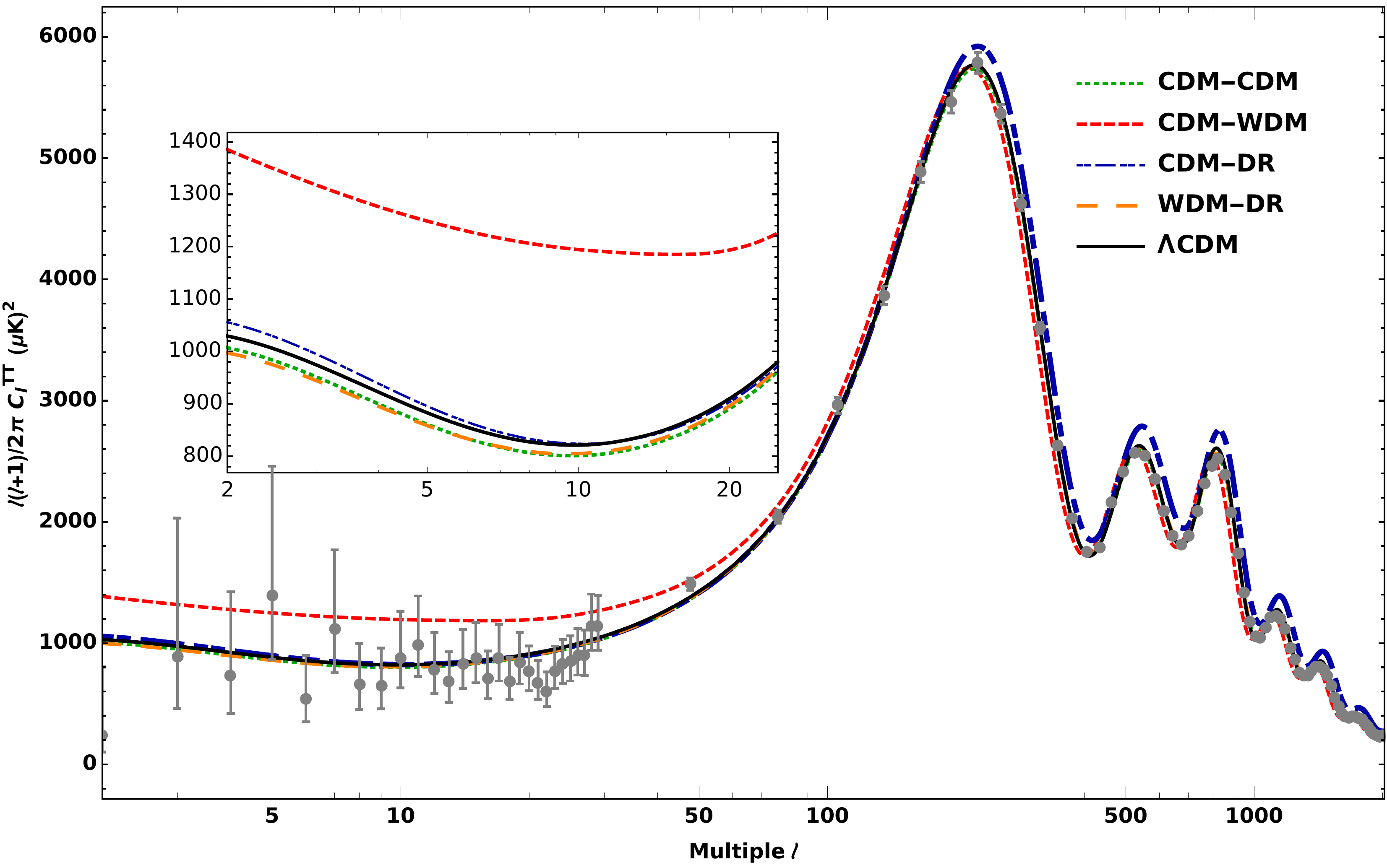}
		\includegraphics[width=8cm]{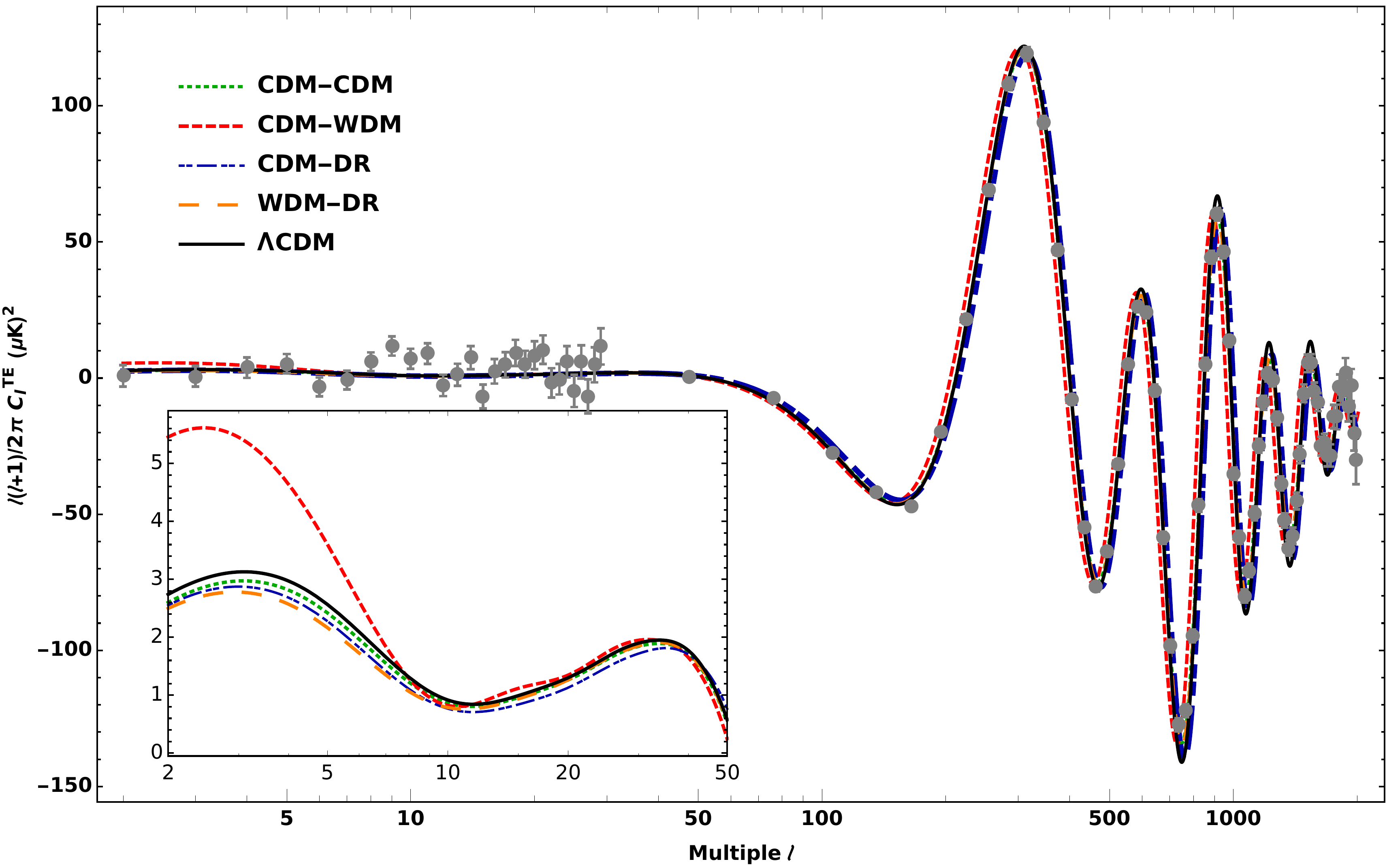}
		\includegraphics[width=8cm]{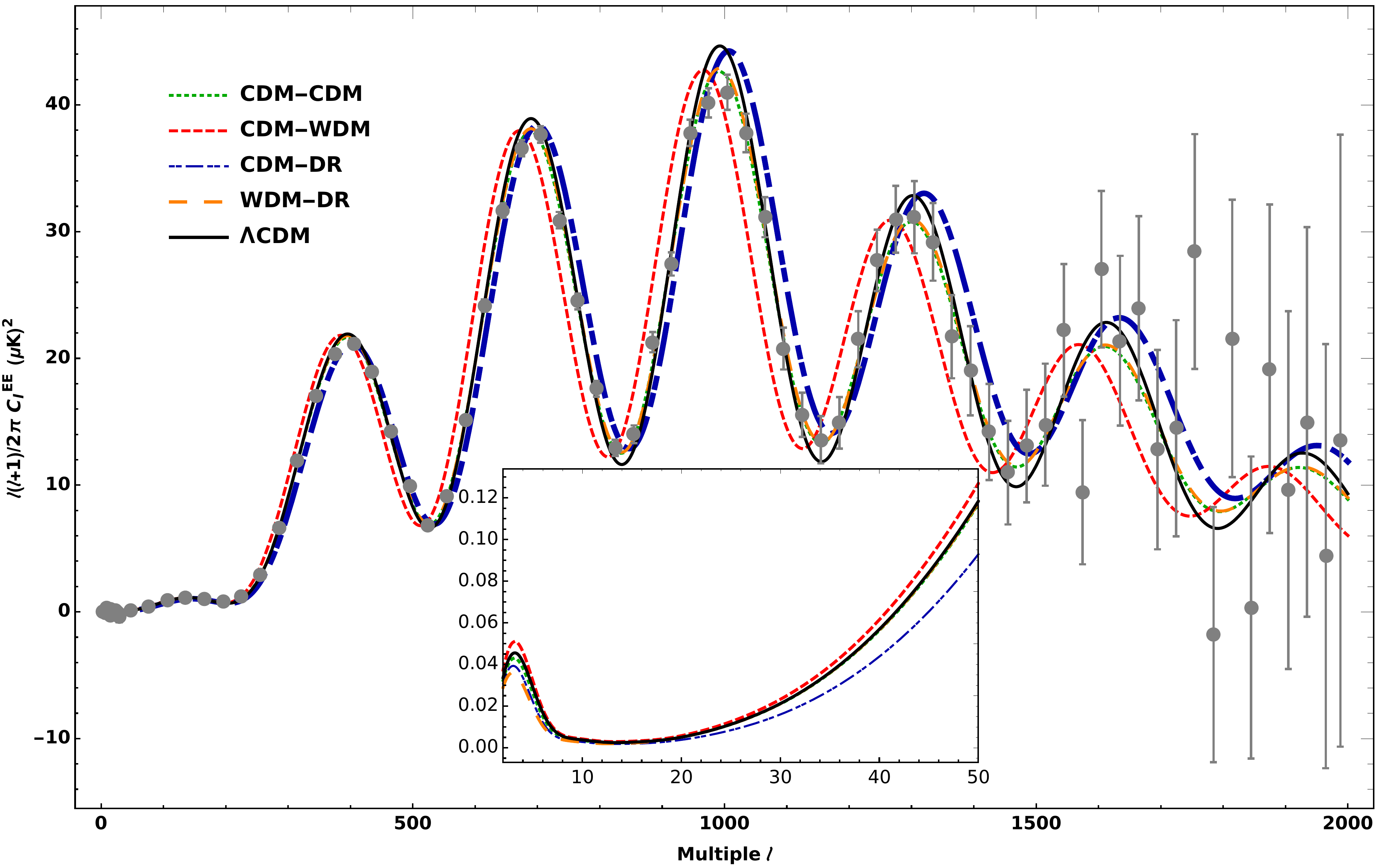}
		\caption{
			Planck 2018 CMB temperature anisotropy power spectrum and
			temperature-polarisation cross-power spectrum TT, TE and EE spectra for the different scenarios in order from top to bottom panel. In the bottom panels, we show the fractional difference between the different models and the $\Lambda$CDM case. In all models, we use the best fit
			values from Tables \ref{tabwdmdr} and \ref{tablcdm}.
		} 
		\label{figcl-tt}
	\end{center}
\end{figure}
In the following, we represent the evolution of the main cosmological quantities based on the best fit values of cosmological parameters are presented in Tables~\ref{tabwdmdr} and~\ref{tablcdm}.  We show the CMB temperature power spectrum TT, TE and EE in order from top to bottom in Figure~\ref{figcl-tt} for the different scenarios. It seems the CDM-WDM case shows a different behavior for the low $\ell$'s and predict a higher value for $C_{\ell}$. In these scales ISW is important and maybe CDM-WDM affects this signal.\\
We plot the evolution of  $H(z)$ in Figure~\ref{figH}. As we see at early times before the onset of decays all decaying models behave identically to the $\Lambda$CDM universe inferred from CMB measurements and at late times, the measurement of $H_0$  is higher than the one obtained from the CMB under $\Lambda$CDM but doesn't alleviate Hubble tension remarkably. Again in this plot H(z) for CDM-WDM is larger than the other cases and maybe it is related to what was mentioned above i.e. the excess in ISW signal.\\
We know the equation of state parameter ($ w = P/\rho $) for CDM and DDM  particles are zero but we can define effective equation of state for DDM  according to $\dot{\rho}_i+3{\cal{H}}(1+w_{\rm eff})\rho_i=0$ as
\begin{equation}
w_{\rm eff, DDM }=\frac{a\Gamma}{3{\cal{H}}}.
\end{equation}
\begin{figure}
	\begin{center}
		\includegraphics[width=8cm]{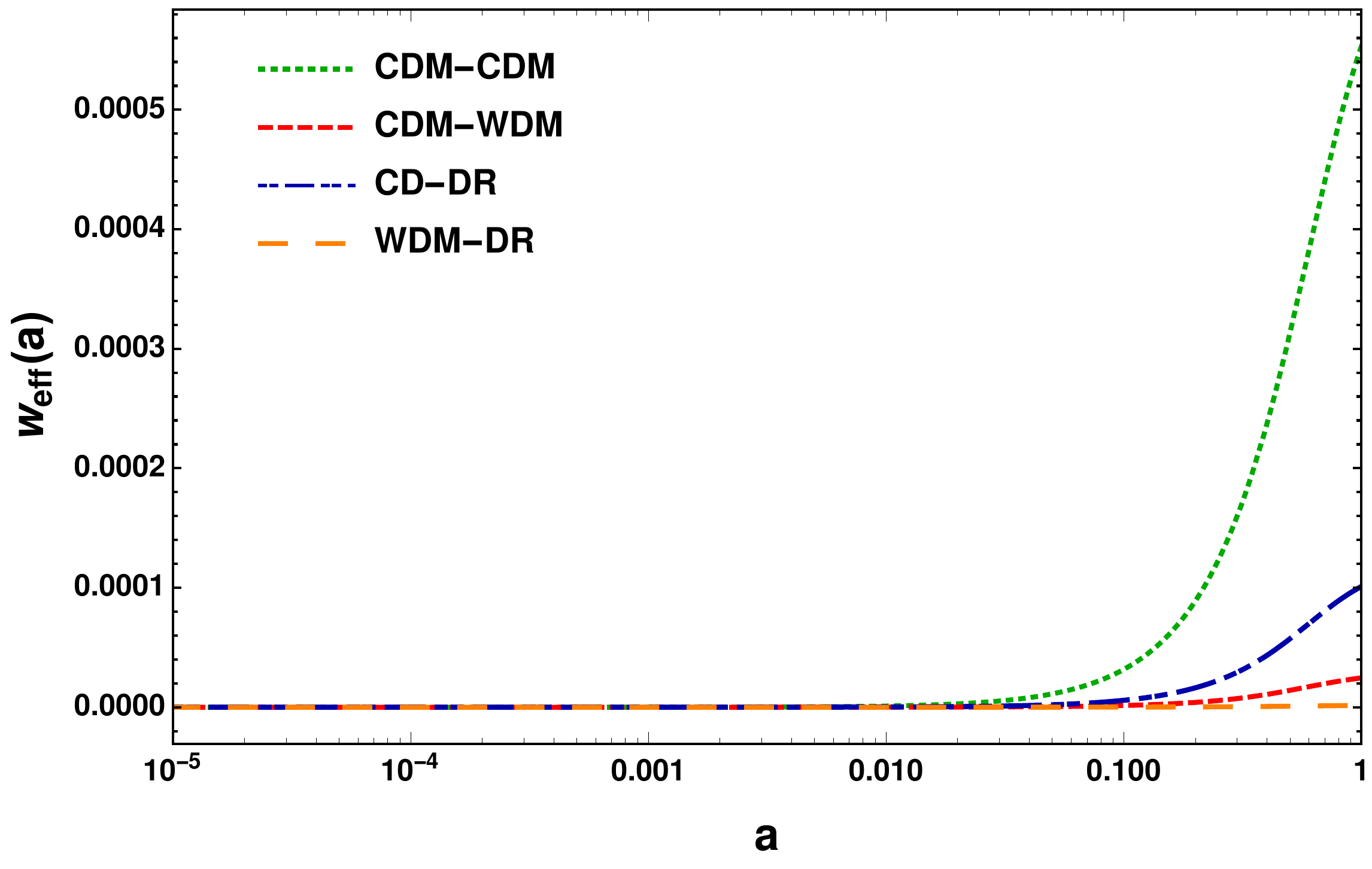}
		
		\caption{The evolution of effective equation of state of decaying dark matter as a function of factor scale. In all models, we use the best fit values from Tables \ref{tabwdmdr} and \ref{tablcdm}.
		} 
		\label{figweff}
	\end{center}
\end{figure} 
\begin{figure}
	\begin{center}
		\includegraphics[width=8cm]{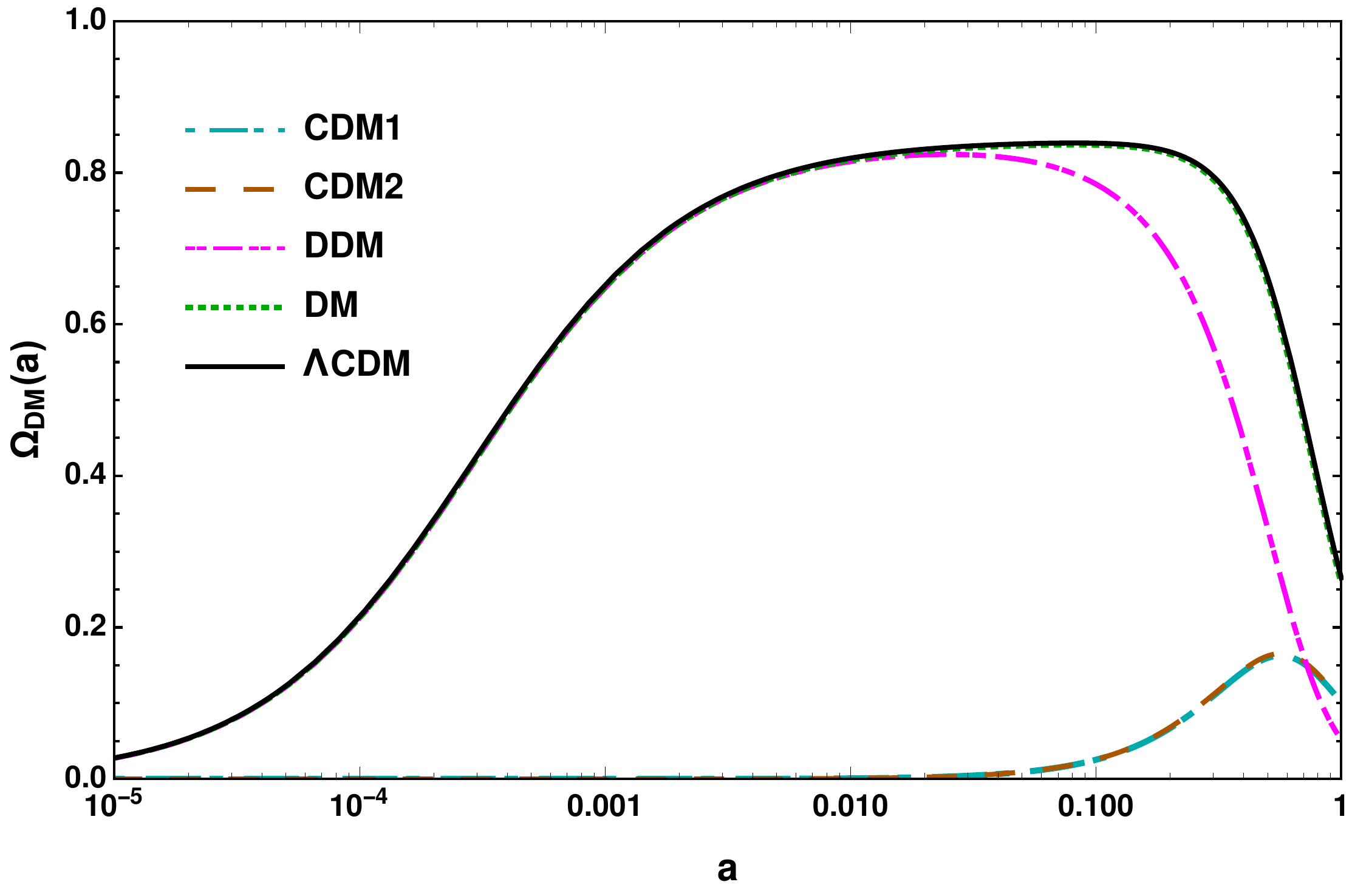}
		\caption{The evolution of the fractional energy density of decaying dark matter in terms of scale factor for CDM-CDM model. We use the best fit
			values from Table~\ref{tabcdm}.
		} 
		\label{figod}
	\end{center}
\end{figure}
In Figure~\ref{figweff} shows the evolution of the effective equation-of-state of mother particle as a function of scale factor, which shows that it is not fully non-relativistic or fully cold and has nonzero value in early time but we observe that it tends to zero at small scale factors (high redshifts) representing the pressureless matter fluid.
At the end of this part, we will mention some important and remarkable results that we have obtained for each model. In Figure~\ref{figod} we have plotted the evolution of the fractional energy densities. In the first scenario, CDM-CDM, for the current time, the contribution of each of the mother and daughter particles are: 0.051
, 0.099 and 0.101, respectively which gives the density ratio $\Omega_{rm DM}/\Omega_B\simeq5.2$ correspond to the amount of cold dark matter in the standard model.\\
In the CDM-WDM scenario, we obtained the proportions of each mother and daughter particle as 0.11, 0.13, and 0.013 for the DDM , CDM, and WDM components, respectively. In this model,  the value of $\Omega_{\rm DM}/\Omega_B$ is  5.5, which is larger than the $\Lambda$CDM model.\\
The third scenario, which has been examined in works such as ~\cite{Blackadder:2015uta,Chudaykin:2016yfk} is the decaying products of non-relativistic particles, CDM and dark radiation, DR. We found the proportions of each mother and daughter particles as 0.222, 0.033 and 0.0001 for the DDM , CDM, and DR components, respectively.\\
In the last studied scenario, WDM-DR, the share of the energy density of the daughter particles is very small and for WDM and DR, respectively equal to 0.003, and 0.0002. 
Some studies, such as~\citep{Abellan:2020pmw} through MCMC comprehensive analysis using up-to-date data in addition to the data we used, including BAO data and the Pantheon SNIa catalog, have shown that the $S_8 $ tension can be resolved if the DDM  experiences this kind of decay due to the suppression in the gravitational clustering induced by WDM free-streaming in a similar fashion to the massive neutrino. Here, we used only Planck data to check if they allow us to add other datasets or not. It seems all the models have problems with higher values of $H_0$ and lower values of $\sigma_8$. This means that these types of models cannot resolve these tensions. This is in agreement with previous results e.g. for the case of daughter DR particles~\citep{Poulin:2016nat,Chudaykin:2016yfk,Bringmann:2018jpr,Anchordoqui:2022gmw,Clark:2020miy}. The only case that seems some hints of solving the $H_0$ tension is the CDM-WDM case. It is obvious that its contour allows for having higher $H_0$ but it makes $\sigma_8$ tension worse. It is worth mentioning that considering the mother particle as warm decaying dark matter (WDDM) component that decays at around the time of matter-radiation equality and they showed that the WDDM can significantly reduce the tension between local and cosmological determinations of the Hubble constant~\cite{Blinov:2020uvz}. \\
 In all scenarios, we obtained the age of the universe to be $\sim$13.8 and the comoving sound horizon found about 147 except for CDM-DR model that obtained $\sim$145. The redshift of matter and radiation equality, $z_{\rm eq}$, of all scenarios was obtained compared to the smaller standard model($z_{\rm eq}<3399$).
\section{Conclusion}\label{concl}
\begin{figure}
	\includegraphics[width=8cm]{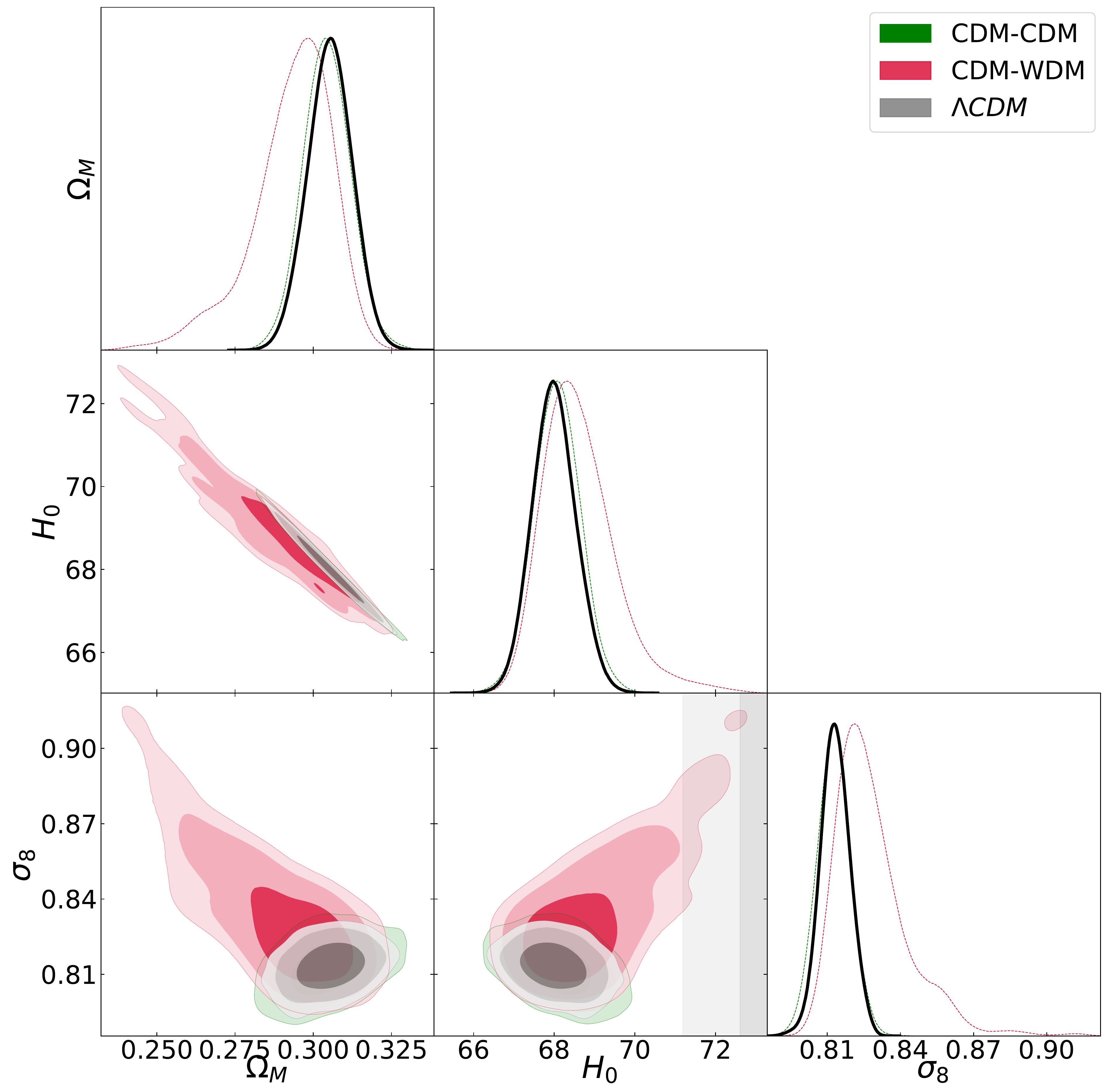}
	\includegraphics[width=8cm]{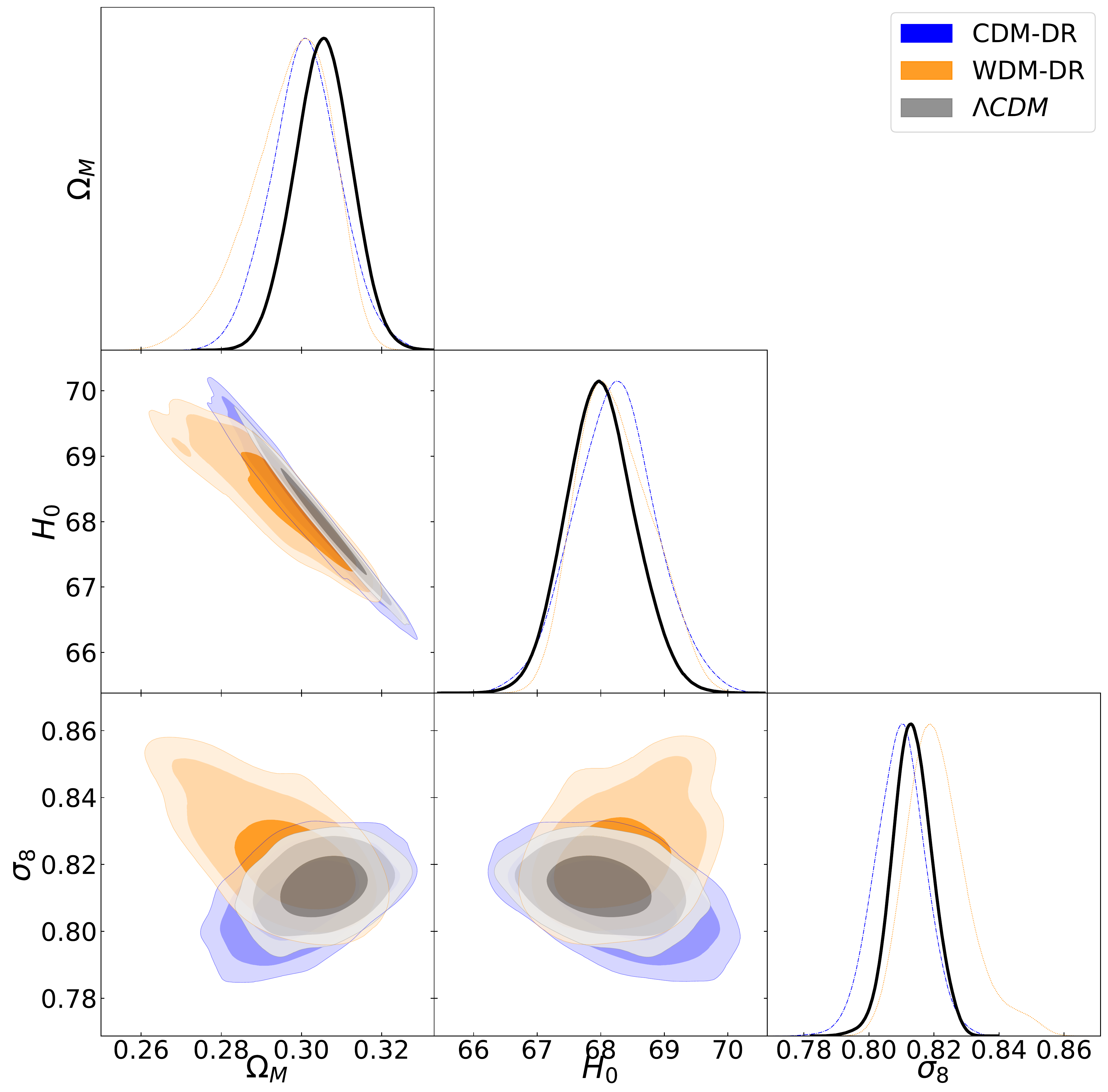}
		\caption{Base $\Lambda$CDM model and decaying scenarios 68\% and 95\% constraint contours on: the matter-density parameter $\Omega_M$ and fluctuation amplitude $\sigma_8$, and the Hubble parameter, $H_0$ using Planck 2018 TT,TE,EE+lowE+lensing.The grey shaded bands refer to $H_0=74.03\pm 1.42$ reported by \citep{Aghanim:2018eyx}.} 
		\label{figcont2}
\end{figure}

In this work, we perform an extensive study on four different cases  of 2-body decaying DM scenarios in the context of solving the $H_0$ and $\sigma_8$ tensions. We discuss in detail their dynamics and their impacts on the CMB anisotropies with a modified version of \texttt{CLASS} and we constrain cosmic parameters and two other degrees of freedom related to decaying scenarios, the decay rate, $\Gamma$, and a fraction of
rest mass energy, $\epsilon$, by performing a Markov Chain Monte Carlo fit on the decaying dark model using \texttt{MontePython} and the
Planck 2018 TT,TE,EE+lowl+lowP+lensing data sets. Based on the results obtained in this study, it seems that these decaying scenarios are not able to resolve these two cosmic tensions (as we see in Figure~\ref{figcont2}). This may mean that we need more complicated dark matter scenarios to address the cosmological tensions or perhaps we should consider the interaction between dark matter and dark energy.

\section{Acknowledgments}
This work has been supported financially by a grant from Basic Sciences Research Fund (No. BSRF-phys-399-06).
\section{Data availability}
No new data were generated or analysed in support of this research.

\bibliographystyle{mnras}
\bibliography{ref}
\end{document}